\documentclass[runningheads]{llncs}

\usepackage[T1]{fontenc}
\usepackage{graphicx}
\usepackage{amssymb}
\usepackage{mathtools}
\usepackage{url}
\usepackage{xurl}
\usepackage{booktabs}
\usepackage{enumitem}

\usepackage[colorlinks=true, allcolors=blue]{hyperref}
\begin{document}
\title{PEB Separation and State Migration: Unmasking the New Frontiers of DeFi AML Evasion}
%
%
\titlerunning{PEB Separation and State Migration}
\author{Yixin Cao\inst{1}\and
Xianfeng Cheng\inst{1}\and
Yijie Liu\inst{2}}
\authorrunning{Y. Cao et al.}
%
\institute{
EigenPhi\\
\email{mars@eigenphi.com} \and
Independent Researcher
}
\maketitle              
\begin{abstract}
Transfer-based anti-money laundering (AML) systems monitor token flows through transaction-graph abstractions, implicitly assuming that economically meaningful value migration is sufficiently encoded in transfer-layer connectivity. In this paper, we demonstrate that this assumption—the bedrock of current industrial forensics—fundamentally collapses in composable smart-contract ecosystems.

We formalize two structural mechanisms that undermine the completeness of transfer-layer attribution. First, we introduce Principal\allowbreak--\allowbreak Execution\allowbreak--\allowbreak Beneficiary (PEB) separation, where intent originators, transaction executors (e.g., MEV searchers), and ultimate beneficiaries are functionally decoupled. Second, we formalize state-mediated value migration, where economic coupling is enforced through invariant-driven contract state transitions (e.g., AMM reserve rebalancing) rather than explicit transfer continuity.

Through a real-world case study of role-separated limit order execution and a constructive cross-pool arbitrage model, we prove that these mechanisms render transfer-layer observation neither attribution-complete nor causally closed. We further argue that simply expanding transfer-layer tracing capabilities fails to resolve the underlying attribution ambiguity inherent in structurally decoupled execution. Under modular composition and open participation markets, these mechanisms are structurally generative, implying heuristic-based flow tracing has reached a formal observational boundary. We advocate for a paradigm shift toward AML based on execution semantics, focusing on the restitution of economic causality from atomic execution logic and state invariants rather than static graph connectivity.

\end{abstract}
\section{Introduction}

\subsection{Transfer-Based AML as a Dominant Abstraction}
Over the past decade, anti-money laundering (AML) practices in cryptoasset ecosystems have largely operationalized monitoring primarily through transfer- or transaction-graph abstractions \cite{weber_bitcoin_2019,bellei_shape_2024}. Industrial analytics platforms—such as Elliptic, Chainalysis, TRM Labs—operationalize AML as a pipeline of transaction tracing, entity clustering, labeling, and risk scoring over these graph structures\cite{elliptic_typologies_2024,chainalysis_crime_2025,trm_crime_2025,mao_regtech_2025}. Regulatory frameworks similarly anchor compliance responsibilities around Virtual Asset Service Providers (VASPs), requiring customer due diligence (CDD), suspicious transaction reporting (STR), and the implementation of the FATF Travel Rule for information exchange \cite{fatf_guidance_2021,fatf_update_2025,eu_reg_2023}.

This paradigm implicitly relies on three structural assumptions common in contemporary transaction tracing literature \cite{kumar_survey_2025,tironsakkul_taint_2019}:
\begin{enumerate}
    \item \textbf{Flow continuity}: Economic value migration is assumed to be isomorphic to the transfer layer, meaning it can be reconstructed from transfer-layer observations via path-based reasoning and graph traversal.
    \item \textbf{Initiator visibility}: It is assumed that suspicious conversions are directly initiated by high-risk addresses or their immediate clusters, maintaining a clear causal link between the actor and the action.
    \item \textbf{Endpoint linkage}: Transfer edges are often treated as reliable proxies for beneficiary attribution or downstream control, assuming that the recipient of a transfer is the ultimate economic beneficiary.
\end{enumerate}

Under these assumptions, illicit activity can be modeled as traceable paths from known tainted sources to downstream recipients or exit points, enabling the use of heuristics like ``poison and hair-cut'' or graph learning-based classification \cite{weber_bitcoin_2019,tironsakkul_taint_2019,victor_address_2020,bellei_shape_2024}. 

However, the rapid evolution of composable decentralized finance (DeFi) has begun to stress-test this abstraction to its breaking point \cite{kitzler_disentangling_2023,bartoletti_composability_2024}. Industry reports have increasingly acknowledged that on-chain observables are often limited to high-level transfer metadata and contract interactions, while critical identity signals, internal bookkeeping, and coordination mechanisms often reside off-chain or in complex intent-matching layers \cite{chainalysis_uk_2023,chainalysis_trends_2024}. As a result, AML systems frequently rely on heuristic thresholds and labeled entity libraries—which recent research indicates are often inconsistent across different commercial platforms \cite{haslhofer2026linking}—rendering them vulnerable to sophisticated, structural adversarial constructions.

\subsection{A Real Case that Violates the Abstraction: Intent-Centric Execution}

Composable smart-contract ecosystems introduce execution patterns in which economic value migration is not necessarily recoverable from transfer-layer observations alone. A representative example is the limit-order-mediated execution model, as implemented by protocols such as 1inch \cite{1inch_docs} and UniswapX. In this intent-centric architecture \cite{chitra_intents_2024}:
\begin{itemize}
    \item \textbf{Off-chain Authorization}: Orders are created and signed off-chain as EIP-712 structured data, meaning the economic intent never appears in the mempool or the transaction history as an initiated call from the Principal.
    \item \textbf{Executor Mediation}: On-chain execution is triggered by independent third-party executors (typically order fillers or MEV bots). From a graph perspective, the executor is the EOA $msg.sender$, not the actual illicit actor.
    \item \textbf{Receiver Decoupling}: The architecture allows the designated receiver of funds to differ from the order maker, facilitating a direct "leap" of value across the ledger without an intervening transfer edge between the two parties.
    \item \textbf{Arbitrary Predicates}: Predicate conditions and pre/post-interaction callbacks allow arbitrary embedded logic, enabling value migration to be contingent on state variables (e.g., specific pool prices) that are invisible to transfer-graph traversal \cite{bartoletti_composability_2024}.
\end{itemize}

Consequently, a significant value conversion may occur without the markers required by traditional AML heuristics: (1) a swap transaction initiated by the high-risk address; (2) a transfer-layer-derivable attribution linking the maker to the receiver; or (3) an observable graph structure that uniquely determines economic causality.

Audit reports from leading security firms \cite{openzeppelin_audit_2024,chainsecurity_assessment_2023} further indicate that even systems designed to mitigate MEV may introduce new layers of opacity where internal settlements (e.g., PMM-to-user flows) bypass the visible token swap events typically indexed by forensic tools. This separation between intent, execution, and benefit realization fundamentally weakens the assumption that transfer edges approximate economic control, identifying a critical structural boundary for current forensic paradigms.

\subsection{From Case to Structural Question}
This observation raises a broader structural question: Is transfer-graph incompleteness merely a corner case of specific protocols, or a systematic limitation in composable execution environments? 

In composable smart-contract ecosystems, open participation and market-driven execution structurally relocate economic coupling from transfer-layer continuity to execution-layer coordination \cite{roughgarden_postmev_2024}. Signed intents may be enforced by independent third parties through modular contract composition and searcher markets. Importantly, such role separation is not confined to a specific protocol design. Modular composition and open execution render \textbf{Principal-Execution-Beneficiary (PEB) separation} a structurally generative feature under protocol composability \cite{kitzler_disentangling_2023}. Similar constructions may arise whenever intent expression, transaction submission, and settlement delivery are decoupled across composable components. 

As routing layers, liquidity sources, and the MEV supply chain evolve~\cite{daian_flash_2019,yang2025decentralization}, distinct execution paths may produce observationally indistinguishable net economic outcomes while differing drastically at the transfer layer. If economic migration is not necessarily recoverable from transfer-layer observation, then transfer-based AML is structurally incomplete as an observational abstraction of execution-level causality.

This leads to a fundamental abstraction gap: In composable systems, economic value migration is enforced at the execution layer, while transfer-based AML observes only its projection onto the transfer layer. When multiple execution paths induce indistinguishable transfer patterns, transfer-layer reasoning alone becomes insufficient to uniquely recover economic causality.
\subsection{Contributions}
Motivated by this structural gap, this paper makes the following contributions:
\begin{enumerate}
    \item We formalize PEB separation in the AML context, defining the functional decoupling of intent originators from transaction executors.
    \item We introduce state-mediated value migration as a structural execution-level mechanism that bypasses transfer-layer connectivity.
    \item We show that these two mechanisms jointly imply transfer-layer incompleteness, proving that even full trace visibility is not guaranteed to eliminate attribution ambiguity.
    \item We demonstrate structural generativity under composability, characterizing a fundamental blind spot in prevailing monitoring abstractions.
\end{enumerate}

\subsection{Paper Overview}
The remainder of this paper proceeds as follows. Section 2 reviews related work in AML tracking and execution analysis. Section 3 presents the limit-order-mediated case study to illustrate structural role separation. Section 4 formalizes the observability model and derives structural non-identifiability propositions. Section 5 constructs a generalized state-mediated arbitrage migration model. Section 6 discusses monitoring implications and observational boundaries. Section 7 concludes.

\section{Related Work}

\subsection{Industrial Graph-Based AML Systems}
Industrial blockchain analytics platforms such as Elliptic, Chainalysis, and TRM Labs provide transaction tracing, clustering, exposure scoring, and cross-chain graph visualization tools \cite{elliptic_typologies_2024,chainalysis_crime_2025,trm_crime_2025}. Their methodologies typically combine multi-hop tracing and taint analysis with address-to-entity clustering and risk scoring workflows. Their methodologies typically combine multi-hop tracing and taint analysis with address-to-entity clustering and risk scoring workflows. These systems rely on curated entity labels and protocol-aware heuristics to interpret transfer-layer observations. While effective for known typologies, they require continual rule updates to handle novel composable structures. Public reports acknowledge that attribution baselines may shift as new intelligence becomes available, reflecting the dynamic nature of on-chain observability \cite{chainalysis_trends_2024,chainalysis_uk_2023}.

\subsection{Regulatory Frameworks}
Regulatory approaches center on VASP-based compliance. FATF Recommendation 15\cite{fatf_recommendations} and its updated guidance extend AML/CFT obligations to VASPs, including CDD and Travel Rule information transmission \cite{fatf_guidance_2021,fatf_update_2025}. EU Regulation 2023/1113 similarly imposes transfer-of-funds information requirements for cryptoassets \cite{eu_reg_2023}. In the United States, regulators emphasize sanctions compliance and the usage of blockchain analytics \cite{ofac_guidance_2021}. However, these frameworks presuppose identifiable intermediaries. In permissionless DeFi environments without custodial entry points, enforcement typically relies on exposure tracking through observable transfers, as execution-level intent semantics are not directly embedded in regulatory reporting pipelines \cite{mao_regtech_2025}.

\subsection{Academic Transfer-Graph Learning}
Academic AML research has predominantly adopted transaction- or transfer-graph abstractions. Weber et al. applied GCNs for illicit transaction classification over the Elliptic dataset \cite{weber_bitcoin_2019}. Subsequent work explored subgraph representation learning to capture laundering structures at scale \cite{bellei_shape_2024} and large-scale tracing systems such as TRacer \cite{wu_tracer_2022}. While these approaches advance classification performance, they fundamentally operate over transfer-layer graph abstractions and labeled datasets. Distribution shift and adversarial adaptation pose persistent challenges, particularly when laundering strategies exploit structural composability rather than mere path obfuscation. Furthermore, taint-analysis literature has highlighted methodological instability: different propagation rules can produce significantly different contaminated path sets \cite{tironsakkul_taint_2019}.

\subsection{Execution-Trace-Based Analysis}
Parallel to AML research, security-oriented program analysis frameworks demonstrate the feasibility of extracting execution semantics from transaction traces. TxSpector replays transactions to obtain bytecode-level traces and applies logic rules to detect anomalous interactions \cite{zhang_txspector_2020}, while MFTracer introduces high-fidelity execution-trace analysis to resolve illicit money flows concealed within complex internal calls \cite{yang_mftracer_2025}. Simultaneously, MEV research has revealed how ordering dependencies create alternative value channels beyond direct transfers \cite{daian_flash_2019,yang_sok_mev_2022,roughgarden_postmev_2024}. These works show that execution semantics are computationally extractable \cite{jiao2020semantic}. However, they primarily target vulnerability detection or MEV characterization rather than AML intent modeling, leaving the mapping between trace semantics and laundering intent underdeveloped.

\subsection{Limit-Order-Mediated Execution Structures}
The complexity of intent-centric execution is exemplified by limit-order protocols, which feature off-chain signed orders, third-party filling, and receiver decoupling. Audit assessments note that even MEV-protected mechanisms may be observable or reordered at alternative layers of the execution stack \cite{openzeppelin_audit_2024,chainsecurity_assessment_2023}. Such architectures demonstrate how economic migration can be encoded in execution semantics without transfer-layer structures that uniquely recover beneficiary attribution. Recent studies on DeFi composition highlight that the interaction between composable role separation and transfer-layer attribution remains an open challenge \cite{kitzler_disentangling_2023,haslhofer2026linking}.

\section{Limit-Order-Mediated Value Migration Under Role Separation}

We analyze an Ethereum Mainnet transaction%
\footnote{hash:\allowbreak\texttt{0x04c43669c930a82f9f6fb31757c722e2c9cb4305eaa16baafce378aa1c09e98e}} (Block 15,937,667) in which approximately 38.2M USDC were converted into 38.1M DAI via a 1inch limit order filled by an independent MEV bot. The execution involved the 1inch Limit Order Protocol, Uniswap V3, and Aave flash liquidity, serving as a concrete instance of \textbf{Principal-Execution-Beneficiary (PEB)} role separation.

\subsection{Role Identification: Principal, Executor, Beneficiary}
We denote the participating entities as follows:
\begin{itemize}
    \item \textbf{P (Principal)}: the phishing-labeled maker address that committed USDC through a signed off-chain limit order.
    \item \textbf{E (Executor)}: the independent MEV bot (Filler) that submitted and initiated the on-chain transaction.
    \item \textbf{B (Beneficiary)}: the address that ultimately received the DAI settlement.
\end{itemize}

In this transaction, the roles are functionally disjoint ($P \cap E \cap B = \emptyset$). The principal signs the intent but does not submit the transaction; the executor initiates the call but retains only arbitrage profit rather than the principal proceeds; and the beneficiary receives the funds without interacting with the swap logic.

\subsection{Structural Role Separation}
The conversion does not occur through a direct swap initiated by the principal's address. Instead, the signed limit order created by $P$ is filled by $E$, and the on-chain transaction is entirely initiated by $E$. From a transaction-layer perspective, $P$ does not appear as the $msg.sender$. The conversion is therefore \textit{third-party-initiated}, establishing a structural separation between the risk-labeled principal and the execution environment.

\subsection{Execution-Level Economic Coupling}
\begin{figure}[htb]
    
	\centering
    
	\includegraphics[width=\linewidth]{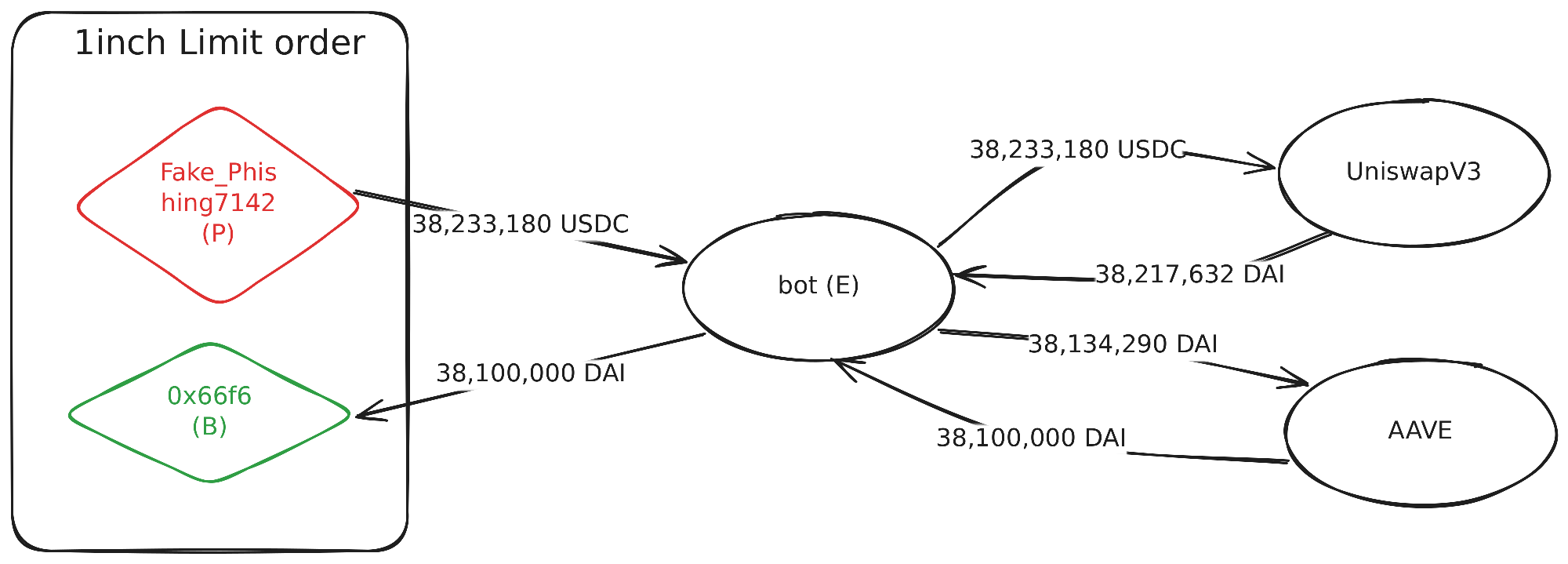}
	\caption{\textbf{Token flow of the USDC-to-DAI conversion via limit-order-mediated MEV execution.} Stolen USDC is committed by the phishing-labeled address through a 1inch limit order and filled by an independent MEV bot, resulting in AMM-settled conversion into DAI delivered to a distinct EOA. No direct token transfer occurs between the maker and the receiver addresses, and no transfer-layer structure uniquely encodes their economic coupling.
}
	  \label{fig_case1}
\end{figure}

Although roles are separated, the economic migration is deterministically enforced within a single atomic transaction. The execution sequence (see Fig.~\ref{fig_case1}) follows a composable pattern:

\begin{enumerate}
    \item $E$ borrows DAI via Aave flash liquidity.
    \item $E$ uses the borrowed DAI to fill the limit order signed by $P$.
    \item $E$ receives the $P$'s USDC from the 1inch settlement contract.
    \item $E$ swaps the USDC for DAI through a Uniswap V3 pool.
    \item $E$ repays the Aave flash loan.
    \item The net proceeds (DAI) are delivered to $B$.
\end{enumerate}

The net economic effect is a migration: $USDC_{at P} \xrightarrow{atomic} DAI_{at B}$. This coupling is achieved through composable protocol logic and AMM state transitions rather than direct transfer continuity. In this competitive searcher market \cite{chitra_intents_2024}, MEV incentives ensure that signed intents are executed efficiently, regardless of the principal's risk profile.

\subsection{Transfer-Layer Attribution Limits}
At the transfer layer, observable edges include $P$ transferring USDC to the bot contract, which finally transfers DAI to $B$. While these edges reflect physical token movements, they do not uniquely encode the principal-beneficiary relationship. Traditional path-finding algorithms cannot deterministically infer that $B$ is the economic recipient of $P$'s asset outflow, as the linkage is mediated through the searcher's coordination and the pool's invariant rebalancing. Consequently, while token flows are visible, beneficiary attribution is not recoverable solely from the transfer-graph abstraction.

\subsection{Structural Generality}
\begin{figure}[htb]
	\centering
	\includegraphics[width=\linewidth]{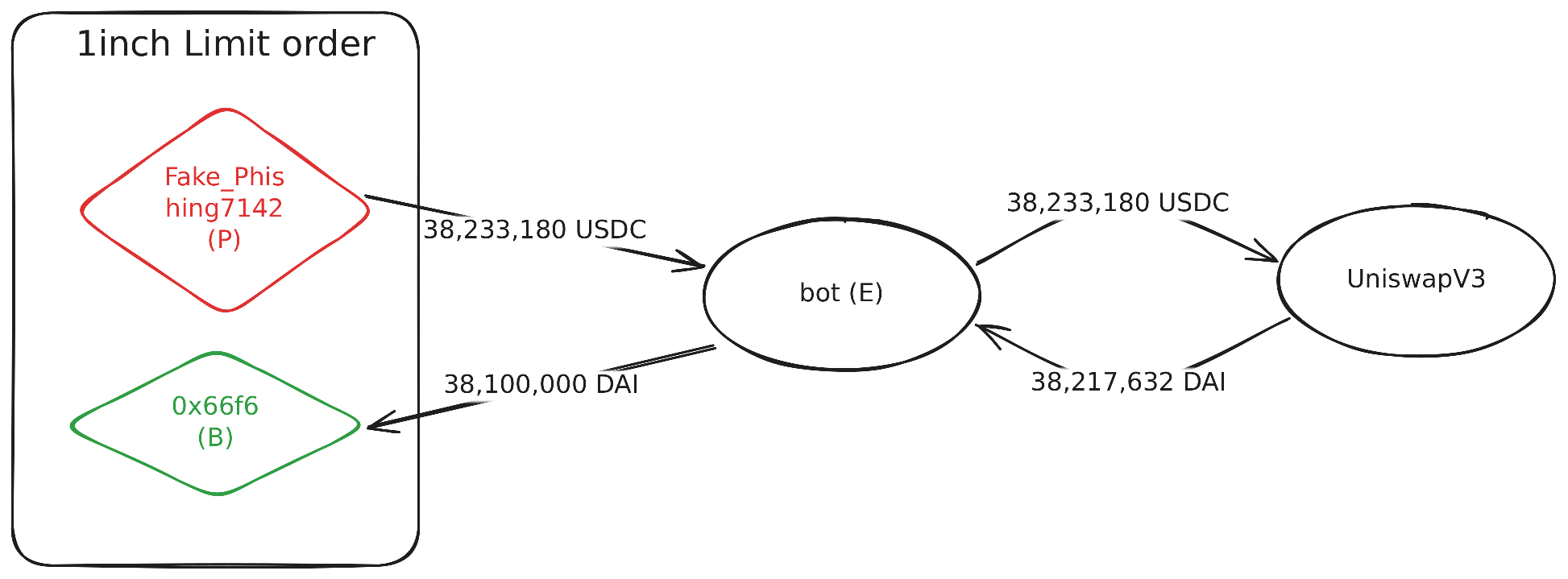}
	\caption{\textbf{Variant using AMM flash swap instead of external flash loan.} The MEV filler sources DAI directly from the AMM via flash swap, fills the 1inch limit order, and repays the pool within the callback. The net state transition is equivalent to the Aave-based execution, demonstrating that external lending is not structurally required.}
	\label{fig_case1_g}
\end{figure}

The use of Aave flash liquidity in this case is incidental; equivalent outcomes are achievable via Uniswap V3 flash swaps with callback repayments (see Fig.~\ref{fig_case1_g}). Thus, the construction is structurally internal to composable execution. As additional routing layers, liquidity sources, and solver strategies are composed, the space of role-separated migration expands combinatorially. Distinct call graphs and settlement paths may yield identical economic outcomes while differing drastically at the transfer layer, rendering pattern-matching detection approaches inherently incomplete.

\section{Failure Model: Transfer-Layer Incompleteness}

\subsection{Transfer-Layer Observation Model}
We model transfer-based AML as an observational abstraction over execution. Let $G = (V, E)$ denote the transfer graph where $V$ is the set of on-chain addresses and $E \subseteq V \times V \times \mathcal{A}$ represents directed token transfer events for assets $\mathcal{A}$. For a fixed asset $A$, we write $G_A = (V, E_A)$. A transfer-layer observer has access to explicit token transfers (edges), address-level interactions, and transaction metadata. However, the observer does not directly access contract internal state variables (e.g., AMM reserves), invariant dynamics, or off-chain intent coordination.

We study whether economic migration can be uniquely reconstructed under this observational model.

\subsection{Economic Migration and Transfer-Layer Recoverability}
We distinguish economic causality from transfer-layer observability.
\begin{definition}[Economic Migration]
Let $P$ (principal) and $B$ (beneficiary) be addresses. An execution instance induces an economic migration of asset $A$ from $P$ to $B$ if:
$$\Delta \text{balance}_A(P) < 0, \quad \Delta \text{balance}_A(B) > 0$$
and these balance changes are jointly enforced by a single atomic execution structure.
\end{definition}

\begin{definition}[Transfer-Layer Recoverability]
An economic migration from $P$ to $B$ is transfer-recoverable if there exists a directed path in $G_A$ that uniquely and deterministically represents the causal linkage between the decrease at $P$ and the increase at $B$.
\end{definition}
Uniqueness is crucial: it requires that the path encodes the specific relationship induced by the execution, rather than being one of many indistinguishable multi-hop routes through shared contracts.

\subsection{Failure Type I: Role-Mediated Indeterminacy}
Under PEB separation, the on-chain transaction realizing value migration is initiated by an address $E \neq P, B$. 
\begin{proposition}[Role-Mediated Indeterminacy]
Under PEB separation, transfer-layer observation may fail to uniquely attribute beneficiary realization to the principal, even when economically enforced migration occurs.
\end{proposition}
The principal may not appear as the initiator, and no uniquely attributable transfer path may link $P$ to $B$. This leads to attribution incompleteness.

\subsection{Failure Type II: State-Mediated Causal Non-Closure}
Composable protocols often encode value redistribution through state transitions governed by invariant functions.
\begin{proposition}[State-Mediated Non-Closure]
Transfer-layer abstraction is not causally closed with respect to economic migration when value relocation is enforced through execution-level state transitions that do not induce a uniquely attributable directed path in the transfer graph.
\end{proposition}
Identical transfer-layer structures may arise under ordinary arbitrage or liquidity rebalancing, making the transfer-layer insufficient to deterministically recover the economic linkage.

\subsection{Unified Transfer-Layer Incompleteness}
The two failure modes above share a structural consequence.
\begin{theorem}[Transfer-Layer Incompleteness, Informal]
In composable smart-contract environments supporting modular execution and open participation, there exist execution patterns for which transfer-layer observations are neither attribution-complete nor causally closed with respect to execution-level economic migration.
\end{theorem}
This incompleteness does not require protocol malfunction; it arises under economically rational execution within standard DeFi architectures.

\subsection{Structural Generativity}
Composable environments exhibit modular composition and open participation under market incentives. Under these conditions, failure Type I and II are not anomalies but structurally generative. As routing layers and execution strategies evolve, distinct execution graphs may yield observationally similar balance outcomes while differing at the transfer layer. Scalability of transfer-based monitoring is thus inherently limited.

\section{State-Mediated Value Relocation: A Constructive Proof}

\subsection{Introduction and Mapping}
Section 4 established that transfer-layer abstraction is neither attribution-complete nor causally closed under composable execution. We now extend this analysis to a more general constructive class: \textbf{State-Mediated Value Migration}. Unlike the limit-order case, which relies on explicit intent delegation, the arbitrage construction embeds economic value relocation within invariant-driven reserve dynamics \cite{qin_arbitrage_2022,heimbach_arbitrage_2024}. 

\subsection{The Construction: Setup and Assumptions}
We formalize a value migration structure implemented within a constant-product Automated Market Maker (AMM) system, which achieves an economic migration without a uniquely attributable path in the transfer graph $G_A$.
\begin{itemize}
    \item \textbf{Assets and Pools}: Consider two independent constant-product pools, $P_1$ and $P_2$, trading the $(A, B)$ pair. Let reserves be $\mathbf{S}_i = (r_A^{(i)}, r_B^{(i)})$, governed by $r_A^{(i)} \cdot r_B^{(i)} = k_i$.
    \item \textbf{Role Mapping}: Let $P$ be the Principal holding asset $A$, $B$ be the Beneficiary receiving asset $A$, and $O$ be the Operator submitting the atomic bundle.
    \item \textbf{Objective.}
Construct an execution such that
\[
\Delta \mathrm{balance}_A(P) = -a,
\qquad
\Delta \mathrm{balance}_A(B) = +a,
\]
without a traceable flow path.
\end{itemize}

\subsection{Mechanism and Execution: The Four-Step Atomic Sequence}
\begin{figure}[htb]
	\centering
	\includegraphics[width=\linewidth]{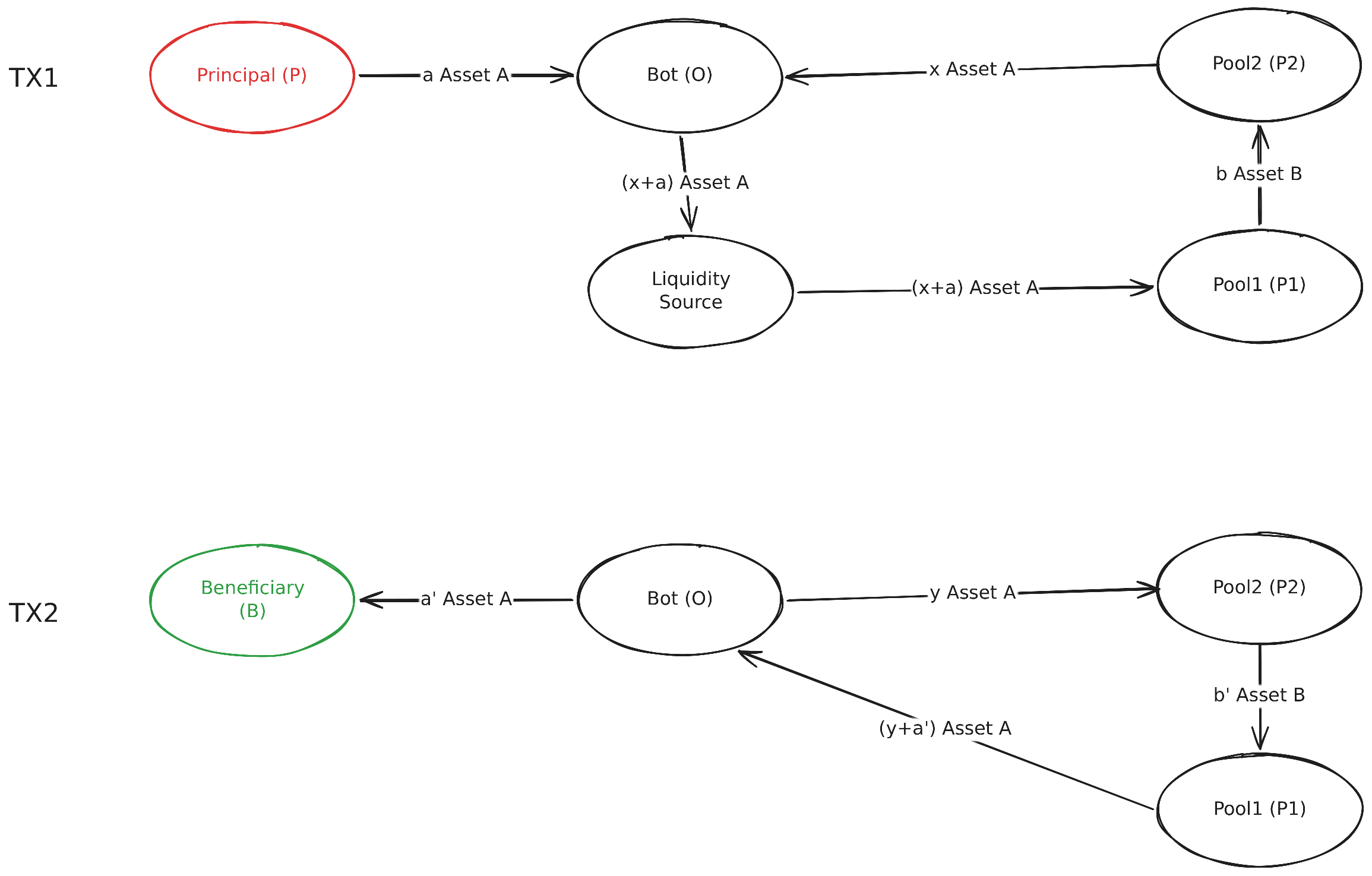}
	\caption{Token flow illustration of constructed transactions}
	\label{fig_case2}
\end{figure}

The construction consists of two back-to-back cross-pool arbitrage loops executed by the Operator $O$ within a single atomic Sequence. This sequence is divided into two distinct logical phases, TX1 and TX2, as illustrated in Fig.~\ref{fig_case2}.

\subsubsection*{Phase 1: TX1 – Dislocation (Steps 1 \& 2)}
In this phase, the operator creates a state imbalance (dislocation) between the pools by injecting capital from the Principal.
\begin{itemize}
    \item \textbf{Step 1 ($P_1, A \to B$)}: $O$ inputs $(a + x)$ units of asset $A$ to $P_1$ via flash liquidity. $O$ receives $b$ units of asset $B$.
     \item \textbf{Step 2 ($P_2, B \to A$)}: $O$ inputs $b$ units of $B$ to $P_2$. The pool updates, and $O$ receives $x$ units of asset $A$, which is used to immediately repay the flash liquidity together with $a$ Asset $A$ which is sourced from $P$ via an allowance.
\end{itemize}

\subsubsection*{Phase 2: TX2 – Extraction (Steps 3 \& 4)}
In this phase, the operator harvests the state dislocation to deliver funds to the Beneficiary.
\begin{itemize}
    \item \textbf{Step 3 ($P_2, A \to B$)}: $O$ flashswaps $b'$ units of asset $B$ from $P_2$.
    \item \textbf{Step 4 ($P_1, B \to A$)}: $O$ inputs $b'$ units of $B$ to $P_1$, receiving $y+a'$ units of asset $A$. $y$ units of asset $A$ is used to repay $P_2$.
\end{itemize}
By selecting parameter $y$ such that $a' = a$, the net economic effect is established: $P$ loses $a$; $B$ gains $a$. Crucially, no direct transfer exists between $P$ and $B$.

\subsubsection*{Proof and Validation}
The structural integrity and feasibility of this mechanism are further detailed in the appendices:
\begin{itemize}
    \item \textbf{Appendix A} provides a formal \textbf{constructive proof} within a theoretical zero-fee constant-product AMM environment, deriving the exact consistency conditions required for the migration.
    \item \textbf{Appendix B} offers \textbf{empirical validation} through an on-chain simulation on an Ethereum Mainnet fork. It demonstrates that even in a \textbf{real-world environment with protocol fees} (e.g., Uniswap V2/SushiSwap 0.3\% fees), an adversary can achieve a relocation efficiency ($\eta$) of approximately 93.5\%.
\end{itemize}

\subsection{Non-Closure and Attribution Ambiguity}
At the transfer layer, the observable edges do not uniquely encode causality. The intermediate transfers form complex loops characteristic of benign market arbitrage. Identical transfer-layer structures can be generated by ordinary routing without intentional value relocation. Since the economic coupling is encoded in the execution-level state transitions and not in the transfer-layer path, the system satisfies \textbf{Proposition 2 (State-Mediated Non-Closure)}.

\subsection{Structural Generativity and Monitoring Implications}
As the number of composable liquidity venues ($n$) and potential routing permutations ($m$) increase, the construction space expands super-linearly. This confirms the premise of \textbf{Theorem 1 (Transfer-Layer Incompleteness)}. The inadequacy of transfer-based AML is a result of a fundamental abstraction gap between the observable transfer layer and execution-level economic causality.

\subsection{Independence from PEB Separation}
The state-mediated construction demonstrates that role separation is sufficient but not necessary to induce transfer-layer incompleteness. Even if the principal is the operator ($O = P$), causal non-closure remains. Thus, the failure mode is structurally internal to composable execution itself.

\section{Discussion: Observational Limits and Monitoring Implications}

\subsection{Transfer-Layer Monitoring as an Incomplete Abstraction}
Sections 3-5 demonstrate that economically enforced asset migration can occur under composable execution without uniquely recoverable principal-beneficiary linkage at the transfer layer. This failure arises because transfer-layer monitoring abstracts execution semantics into edge connectivity. In composable ecosystems, economic causality is encoded in dimensions beyond transfer continuity: (i) off-chain intent coordination (PEB separation); (ii) invariant-driven reserve dynamics ($\mathbf{S}_i$ updates); and (iii) cross-contract call graphs. These dimensions are not preserved under transfer-layer projection, rendering the $\mathbf{G} = (V, E)$ abstraction structurally incomplete.

\subsection{Persistent Asymmetry in Detection}
The structural generativity of these failure modes creates an inherent asymmetry akin to polymorphic or metamorphic malware in computer security. Just as advanced malware can mutate its binary signature while retaining core functionality, adversarial constructions can mutate their observable transfer-layer structure while achieving identical net economic migration. As routing layers and MEV supply chains evolve, the space of possible execution graphs expands combinatorially, outpacing rule-based or motif-based detection space.

\subsection{Limitations of Graph-Based AML Methods}
Existing graph-learning approaches operation over $\mathbf{G}_A$ are bounded by this incompleteness. Even with perfect classification of transfer motifs, structural ambiguity persists as the same motifs may correspond to benign arbitrage or illicit relocation. Rapid innovation introduces persistent distribution shift, rendering models trained on historical transfer-layer patterns increasingly fragile.

\subsection{Towards Execution-Semantic Monitoring}
The identified structural boundary necessitates a shift toward execution-semantic monitoring, mirroring the evolution of security from signature-based to behavioral analysis. This requires explicit reasoning about call graphs and internal state transitions \cite{jiao2020semantic,yang_mftracer_2025}. However, this shift introduces new challenges: (1) \textit{Attribution Ambiguity}, as intents must be disentangled from mechanical execution; and (2) \textit{Economic Intent Inference}, requiring models for coordination and off-chain signaling \cite{haslhofer2026linking}.

\subsection{Defining the Structural Boundary}
We conclude that transfer-layer monitoring is structurally insufficient in environments exhibiting concurrent modular composition, open participation under market incentives, and invariant-driven state dynamics. This work establishes a formal boundary for prevailing AML abstractions, advocating for a focus on execution-level causality modeling.

\section{Conclusion}
This work identifies two orthogonal structural mechanisms--PEB separation and state-mediated value migration--that jointly define the limit of transfer-based AML. We show that in composable DeFi environments, the dominant abstraction of transaction tracing fails to uniquely recover economic migration. These failures emerge from economically rational execution within standard protocol architectures, rendering them structurally generative. Future research must formalize the integration between execution semantics and AML attribution to ensure compliance in an increasingly modular Web3 landscape.


%
%
%
\bibliographystyle{splncs04}
\bibliography{mybibliography}

@misc{chainalysis_crime_2025,
  author = {Chainalysis},
  title = {The 2025 Crypto Crime Report},
  year = {2025},
  note = {Industrial Whitepaper, February}
}

@misc{trm_crime_2025,
  author = {TRM Labs},
  title = {2025 Crypto Crime Report},
  year = {2025},
  note = {Industrial Whitepaper}
}

@misc{chainalysis_trends_2024,
  author       = {{Chainalysis}},
  title        = {{Money Laundering and Cryptocurrency: Trends and New Techniques for Detection and Investigation}},
  howpublished = {Chainalysis Blog},
  year         = {2024},
  month        = {July}
}

@techreport{chainalysis_uk_2023,
  author      = {{Chainalysis}},
  title       = {{Written evidence submitted by Chainalysis (FRA0028)}},
  institution = {UK Parliament, Fraud Act 2006 and Digital Fraud Committee},
  type        = {Written Evidence},
  year        = {2023}
}

@misc{elliptic_typologies_2024,
  author = {Elliptic},
  title = {Typologies Report 2024: The Evolution of Illicit Finance in DeFi},
  year = {2024},
  note = {Industrial Whitepaper}
}

@techreport{fatf_recommendations,
  author      = {{Financial Action Task Force (FATF)}},
  title       = {{International Standards on Combating Money Laundering and the Financing of Terrorism and Proliferation - The FATF Recommendations}},
  institution = {FATF},
  year        = {2012}
}

@techreport{fatf_guidance_2021,
  author      = {{Financial Action Task Force (FATF)}},
  title       = {{Updated Guidance for a Risk-Based Approach to Virtual Assets and Virtual Asset Service Providers}},
  institution = {FATF},
  year        = {2021}
}

@techreport{fatf_update_2025,
  author      = {{Financial Action Task Force (FATF)}},
  title       = {{Targeted Update on Implementation of the FATF Standards on Virtual Assets and VASPs}},
  institution = {FATF},
  year        = {2025}
}

@misc{eu_reg_2023,
  author       = {{European Parliament and Council of the European Union}},
  title        = {{Regulation (EU) 2023/1113 on information accompanying transfers of funds and certain crypto-assets}},
  howpublished = {Official Journal of the European Union, L 150/1},
  year         = {2023},
  month        = {June}
}

@techreport{ofac_guidance_2021,
  author      = {{Office of Foreign Assets Control (OFAC)}},
  title       = {{Sanctions Compliance Guidance for the Virtual Currency Industry}},
  institution = {U.S. Department of the Treasury},
  year        = {2021},
  month       = {October}
}

@article{mao_regtech_2025,
  title={SoK: Web3 RegTech for Cryptocurrency VASP AML/CFT Compliance},
  author={Mao, Qian'ang and Wang, Jiaxin and Liu, Ya and Zhu, Li and Chen, Jiaman and Yan, Jiaqi},
  journal={arXiv preprint arXiv:2512.24888},
  year={2025}
}

@article{kumar_survey_2025,
  title={A Survey of Transaction Tracing Techniques for Blockchain Systems},
  author={Kumar, Ayush and Thing, Vrizlynn LL},
  journal={arXiv preprint arXiv:2510.09624},
  year={2025}
}

@inproceedings{haslhofer2026linking,
  title={Linking Cryptoasset Attribution Tags to Knowledge Graph Entities: An LLM-Based Approach},
  author={Haslhofer, Bernhard},
  booktitle={Financial Cryptography and Data Security: 29th International Conference, FC 2025, Miyakojima, Japan, April 14--18, 2025, Revised Selected Papers, Part II},
  pages={366},
  year={2026},
  organization={Springer Nature}
}

@article{weber_bitcoin_2019,
  title={Anti-money laundering in bitcoin: Experimenting with graph convolutional networks for financial forensics},
  author={Weber, Mark and Domeniconi, Giacomo and Chen, Jie and Weidele, Daniel Karl I and Bellei, Claudio and Robinson, Tom and Leiserson, Charles E},
  journal={arXiv preprint arXiv:1908.02591},
  year={2019}
}

@article{bellei_shape_2024,
  title={The shape of money laundering: Subgraph representation learning on the blockchain with the elliptic2 dataset},
  author={Bellei, Claudio and Xu, Muhua and Phillips, Ross and Robinson, Tom and Weber, Mark and Kaler, Tim and Leiserson, Charles E and Chen, Jie and others},
  journal={arXiv preprint arXiv:2404.19109},
  year={2024}
}

@article{tironsakkul_taint_2019,
  title={Probing the mystery of cryptocurrency theft: an investigation into methods for taint analysis},
  author={Tironsakkul, Tin and Maarek, Manuel and Eross, Andrea and Just, Mike},
  journal={arXiv preprint arXiv:1906.05754},
  year={2019}
}

@inproceedings{bartoletti_composability_2024,
  title={DeFi composability as MEV non-interference},
  author={Bartoletti, Massimo and Marchesin, Riccardo and Zunino, Roberto},
  booktitle={International Conference on Financial Cryptography and Data Security},
  pages={369--387},
  year={2024},
  organization={Springer}
}

@article{chitra_intents_2024,
  title={An analysis of intent-based markets},
  author={Chitra, Tarun and Kulkarni, Kshitij and Pai, Mallesh and Diamandis, Theo},
  journal={arXiv preprint arXiv:2403.02525},
  year={2024}
}

@inproceedings{roughgarden_postmev_2024,
  title={Transaction fee mechanism design in a post-mev world},
  author={Bahrani, Maryam and Garimidi, Pranav and Roughgarden, Tim},
  booktitle={6th Conference on Advances in Financial Technologies (AFT 2024)},
  pages={29--1},
  year={2024},
  organization={Schloss Dagstuhl--Leibniz-Zentrum f{\"u}r Informatik}
}

@inproceedings{daian_flash_2019,
  title={Flash boys 2.0: Frontrunning in decentralized exchanges, miner extractable value, and consensus instability},
  author={Daian, Philip and Goldfeder, Steven and Kell, Tyler and Li, Yunqi and Zhao, Xueyuan and Bentov, Iddo and Breidenbach, Lorenz and Juels, Ari},
  booktitle={2020 IEEE symposium on security and privacy (SP)},
  pages={910--927},
  year={2020},
  organization={IEEE}
}

@inproceedings{yang2025decentralization,
  title={Decentralization of ethereum's builder market},
  author={Yang, Sen and Nayak, Kartik and Zhang, Fan},
  booktitle={2025 IEEE Symposium on Security and Privacy (SP)},
  pages={1512--1530},
  year={2025},
  organization={IEEE}
}

@article{kitzler_disentangling_2023,
  title={Disentangling decentralized finance (DeFi) compositions},
  author={Kitzler, Stefan and Victor, Friedhelm and Saggese, Pietro and Haslhofer, Bernhard},
  journal={ACM Transactions on the Web},
  volume={17},
  number={2},
  pages={1--26},
  year={2023},
  publisher={ACM New York, NY}
}

@inproceedings{heimbach_arbitrage_2024,
  title={Non-atomic arbitrage in decentralized finance},
  author={Heimbach, Lioba and Pahari, Vabuk and Schertenleib, Eric},
  booktitle={2024 IEEE Symposium on Security and Privacy (SP)},
  pages={3866--3884},
  year={2024},
  organization={IEEE}
}

@inproceedings{qin_arbitrage_2022,
  title={Quantifying blockchain extractable value: How dark is the forest?},
  author={Qin, Kaihua and Zhou, Liyi and Gervais, Arthur},
  booktitle={2022 IEEE Symposium on Security and Privacy (SP)},
  pages={198--214},
  year={2022},
  organization={IEEE}
}

@article{yang_mftracer_2025,
  title={Shedding light on shadows: Automatically tracing illicit money flows on EVM-compatible blockchains},
  author={Huo, Yicheng and Hu, Yufeng and Zhou, Yajin and Yu, Ting and Wu, Lei and Wang, Cong},
  journal={Proceedings of the ACM on Measurement and Analysis of Computing Systems},
  volume={9},
  number={3},
  pages={1--35},
  year={2025},
  publisher={ACM New York, NY, USA}
}

@inproceedings{jiao2020semantic,
  title={Semantic understanding of smart contracts: Executable operational semantics of solidity},
  author={Jiao, Jiao and Kan, Shuanglong and Lin, Shang-Wei and Sanan, David and Liu, Yang and Sun, Jun},
  booktitle={2020 IEEE Symposium on Security and Privacy (SP)},
  pages={1695--1712},
  year={2020},
  organization={IEEE}
}

@article{wu_tracer_2022,
  title={TRacer: Scalable graph-based transaction tracing for account-based blockchain trading systems},
  author={Wu, Zhiying and Liu, Jieli and Wu, Jiajing and Zheng, Zibin and Chen, Ting},
  journal={IEEE Transactions on Information Forensics and Security},
  volume={18},
  pages={2609--2621},
  year={2023},
  publisher={IEEE}
}

@inproceedings{zhang_txspector_2020,
  title={$\{$TXSPECTOR$\}$: Uncovering attacks in ethereum from transactions},
  author={Zhang, Mengya and Zhang, Xiaokuan and Zhang, Yinqian and Lin, Zhiqiang},
  booktitle={29th USENIX Security Symposium (USENIX Security 20)},
  pages={2775--2792},
  year={2020}
}

@misc{1inch_docs,
  author = {{1inch Network}},
  title = {Limit Order Protocol Documentation},
  howpublished = {\url{https://github.com/1inch/limit-order-protocol}}
}

@techreport{chainsecurity_assessment_2023,
  title       = {Code Assessment of the Limit Order Settlement Smart Contracts},
  author      = {{ChainSecurity}},
  year        = {2023},
  institution = {ChainSecurity},
  type        = {Audit Report},
  note        = {Accessed: 2026}
}

@techreport{openzeppelin_audit_2024,
  title       = {1inch Settlement Refactor Audit},
  author      = {{OpenZeppelin}},
  year        = {2024},
  institution = {OpenZeppelin},
  type        = {Security Audit},
  month       = {May},
  note        = {Accessed: 2026}
}

@inproceedings{yang_sok_mev_2022,
  title={SoK: MEV countermeasures},
  author={Yang, Sen and Zhang, Fan and Huang, Ken and Chen, Xi and Yang, Youwei and Zhu, Feng},
  booktitle={Proceedings of the workshop on decentralized finance and security},
  pages={21--30},
  year={2024}
}

@inproceedings{victor_address_2020,
  title={Address clustering heuristics for Ethereum},
  author={Victor, Friedhelm},
  booktitle={International conference on financial cryptography and data security},
  pages={617--633},
  year={2020},
  organization={Springer}
}

\newpage

\appendix
\section{A Proof of State-Mediated Value Relocation}

We provide a constructive proof demonstrating that in a zero-fee constant-product AMM environment, an economic migration can be enforced through execution-level state transitions without a uniquely attributable transfer path between the principal and beneficiary.

\subsection{Construction: Setup and Assumptions}
Let $G_A$ be the transfer graph for asset $A$. Consider two independent constant-product pools, $P_1$ and $P_2$, trading the same $(A, B)$ asset pair. The state of each pool $i$ is defined by its reserves $\mathbf{S}_i = (r_A^{(i)}, r_B^{(i)})$, governed by the invariant:
\begin{equation}
    (r_A^{(i)})(r_B^{(i)}) = k_i
\end{equation}
We assume zero transaction fees to ensure full reversibility. The participating addresses are:
\begin{itemize}
    \item $P$ (Principal): Holds the initial source asset $A$.
    \item $B$ (Beneficiary): Ultimately receives the target asset $A$.
    \item $O$ (Operator): Executes the atomic transaction bundle.
\end{itemize}

\subsection{The Four-Step Atomic Execution Sequence}
The operator $O$ executes two back-to-back cross-pool arbitrage loops within a single atomic transaction. Let $a$ be the units migrating from $P$, and $x$ be the units sourced via flash liquidity.

\begin{enumerate}
    \item \textbf{Step 1 ($P_1, A \to B$)}: $O$ inputs $(a + x)$ units of $A$ into $P_1$. Amount $a$ is sourced from $P$ via an allowance, and $x$ from flash liquidity.
    \begin{equation}
        (r_A^{(1)} + x + a)(r_B^{(1)} - b) = r_A^{(1)} r_B^{(1)} \implies b = \frac{r_B^{(1)}(x+a)}{r_A^{(1)}+x+a}
    \end{equation}
    
    \item \textbf{Step 2 ($P_2, B \to A$)}: $O$ inputs $b$ units of $B$ into $P_2$, receiving $x$ units of $A$ to repay the flash liquidity.
    \begin{equation}
        (r_B^{(2)} + b)(r_A^{(2)} - x) = r_A^{(2)} r_B^{(2)} \implies b = \frac{r_B^{(2)} x}{r_A^{(2)}-x}
    \end{equation}
    Combining (2) and (3) yields the consistency condition for the first loop:
    \begin{equation}
        r_B^{(1)}(x+a)(r_A^{(2)}-x) = r_B^{(2)} x(r_A^{(1)}+x+a)
    \end{equation}

    \item \textbf{Step 3 ($P_2, A \to B$)}: $O$ inputs $y$ units of $A$ into $P_2$. The state is updated, and $O$ receives $b'$ units of $B$. Using the simplified invariant:
    \begin{equation}
        b' = \frac{r_A^{(2)} r_B^{(2)} y}{(r_A^{(2)} - x)(r_A^{(2)} - x + y)}
    \end{equation}

    \item \textbf{Step 4 ($P_1, B \to A$)}: $O$ inputs $b'$ units of $B$ into $P_1$, receiving $a'$ units of $A$. The parameter $y$ is selected such that $a' = a$.
    \begin{equation}
        y + a' = \frac{(r_A^{(1)} + x + a) b'}{r_B^{(1)} - b + b'}
    \end{equation}
\end{enumerate}

\subsection{Conclusion of Proof}
Setting $a' = a$ in Equation (6) establishes the net economic effect: $\Delta \text{balance}_A(P) = -a$ and $\Delta \text{balance}_A(B) = +a$. Crucially:
\begin{itemize}
    \item \textbf{Attribution Ambiguity}: The transfer edges $P \to  O \to P_1$ and $P_1 \to O \to B$ are separated by arbitrage loops that are observationally indistinguishable from benign market activity.
    \item \textbf{Causal Non-Closure}: The coupling is enforced via coordinated updates to $\mathbf{S}_1$ and $\mathbf{S}_2$ rather than flow continuity.
\end{itemize}
Thus, state-mediated value relocation is structurally sufficient to induce transfer-layer incompleteness, independent of explicit role separation.

\section{Empirical Validation via On-Chain Simulation}

To evaluate the practical feasibility of state-mediated value relocation under realistic market conditions (including protocol fees), we conducted an on-chain simulation using a mainnet fork.

\subsection{Simulation Environment and Parameters}
The simulation was executed in a local \texttt{Anvil} fork of the Ethereum Mainnet at \textbf{Block 21808947}. We utilized two major decentralized liquidity venues:
\begin{itemize}
    \item \textbf{Pool 1 (Uniswap V2)}: WETH/USDT (\texttt{0x0d4a...852})\footnote{pool address:\texttt{0x0d4a11d5EEaaC28EC3F61d100daF4d40471f1852}}
    \item \textbf{Pool 2 (SushiSwap)}: WETH/USDT (\texttt{0x06da...553})\footnote{pool address:\texttt{0x06da0fd433C1A5d7a4faa01111c044910A184553}}
\end{itemize}

The simulation aimed to "wash" a principal amount $a = 10$ WETH from EOA P to EOA B. To amplify the market dislocation, a flash loan of $x = 50.49$ WETH was utilized. Both protocols impose a standard $0.3\%$ swap fee.

\subsection{Quantitative Execution Results}
The simulation successfully achieved the value relocation with high efficiency despite the cumulative impact of fees across four swap steps. Table \ref{tab:sim_results} summarizes the key parameters.

\begin{table}[h]
\centering
\caption{Simulation results for a 10 WETH state-mediated migration.}
\label{tab:sim_results}
\begin{tabular}{@{}lll@{}}
\toprule
\textbf{Parameter} & \textbf{Description} & \textbf{Value (Simulated)} \\ \midrule
$a$ & Input Capital (WETH from EOA P) & 10.0000 \\
$x$ & Flash Loan Amount (WETH) & 50.4893 \\
$b$ & Stage 1 Intermediate (USDT Output) & 159,461.05 \\
$x'$ & Recovered Flash Capital (WETH) & 50.4741 \\
$b'$ & Stage 2 Extraction Input (USDT) & 157,262.60 \\
$y$ & Extraction Repayment (WETH) & 49.9515 \\
$a'$ & Net "Cleaned" Output (WETH to EOA B) & 9.3541 \\ \bottomrule
\end{tabular}
\end{table}

\subsection{Analysis of Efficiency and Fee Loss}
The total efficiency of the relocation is defined as $\eta = a' / a$. In this simulation:
$$\eta = \frac{9.3541}{10.0000} \approx 93.5\%$$
The total loss of $6.5\%$ (0.6459 WETH) represents the "cost of washing." This loss is primarily composed of:
\begin{enumerate}
    \item \textbf{Protocol Fees}: Each of the four swaps incurs a $0.3\%$ fee, totaling $\approx 1.2\%$ in nominal fees.
    \item \textbf{Slippage and Imbalance}: The delta between the dislocation depth in Stage 1 and the extraction volume in Stage 2.
\end{enumerate}

\subsection{Forensic Implications}
The simulation confirms that even with standard LP fees, an adversary can relocate significant economic value with $>90\%$ efficiency. From a transfer-layer perspective, the flows are recorded as typical arbitrage loops ($EOA \to P_1 \to P_2 \to EOA$). No forensic tool restricted to transfer connectivity can uniquely link EOA P's 10 WETH outflow to EOA B's 9.35 WETH inflow, as the causality is entirely contained within the state updates of Pool 1 and Pool 2.

\end{document}